# Bridging the Gap Between Consciousness and Matter: Recurrent Out-of-Body Projection of Visual Awareness Revealed by the Law of Non-Identity

Jinsong Meng[1]*

[1]University of Electronic Science and Technology of China, Chengdu 611731, China
*Correspondence author. Email: mengjinsong@uestc.edu.cn

**Consciousness is an explicit outcome of brain activity. However, the link between consciousness and the material world remains to be explored. We applied a new logic tool, the non-identity law, to the analysis of the visual dynamics related to the naturalistic observation of a night-shot still life. We show that visual awareness possesses a postponed, recurrent out-of-body projection pathway and that the out-of-body projection is superimposed onto the original, which is reciprocally verified by vision and touch. This suggests that the visual system instinctually not only represents the subjective image (brain-generated imagery) but also projects the image back onto the original or to a specific place according to the cues of the manipulated afferent messenger light signaling pathway. This finding provides a foundation for understanding the subjectivity and intentionality of consciousness from the perspective of visual awareness and the isomorphic relationship between an unknowable original, private experience, and shareable expression. The result paves the way for scientific research on consciousness and facilitates the integration of humanities and natural science.**

## 1 Introduction

Consciousness, which is closely related to sensing, emotion, and thinking, is the focus of the mind–body problem long argued by philosophers from different philosophical schools, such as animism, dualism, materialism, idealism, and transcendental philosophy (Kant, 1783). To date, however, it has remained difficult to define consciousness quantitatively. Contemporary philosophers have argued that "consciousness is just about the last surviving mystery" (Dennett, 1991, pp. 21) and that "consciousness is the biggest mystery" and "may be the largest outstanding obstacle in our quest for a scientific understanding of the universe" (Chalmers, 1996, pp. xi).

Due to the close relationship between neuroscience and consciousness, neurobiologists were among the first to initiate scientific research on consciousness (Eccles, 1951; Edelman, 1989; Crick and Koch, 1990, 2003; Koch 2004). There are four closely related questions that require an answer: 1) how does unconscious neural activity respond to an afferent sensory stimulus and memory recall, 2) where does consciousness arise, 3) how does consciousness arise, and 4) what is consciousness? Significant progress (such as in the cell and molecular biology of the neuron, synaptic transmission, and the neural basis of cognition, perception, and movement) has been made regarding the first question over the past century (Albright et al., 2000; Kandel et al., 2012). Regarding the second question, scientific research on the biological substrate of consciousness has identified the most likely neural correlates of consciousness (NCC) — the minimum neural mechanisms jointly sufficient for any one specific conscious experience (Koch et al., 2016). There is, however, a debate as to whether the NCC is local or global (Mashour, 2018). Although somewhat controversial at present, these two questions can be experimentally determined. Contrastingly, the biggest challenge is to explain the conscious experience, i.e., to answer the last two questions mentioned above, termed as the "hard problem" of consciousness (Chalmers, 1995). This "hard problem" has drawn more scientists, including physicists, into this field. As a result, several theories of consciousness, such as the global workspace theory, integrated information theory, and Orch-OR theory have been proposed, each of which, in their own way, explain some aspects of consciousness (Van Gulick, 2018). These theories, however, are neither derived from each other nor provide an adequate explanation of the nature of consciousness on their own. In turn, it is difficult to prove any of these theories unequivocally, thereby giving rise to some controversies (Michael, 2015; Koch and Hepp, 2006). Chalmers argued that there are systematic reasons for the failure of the conventional methods of cognitive science, neuroscience, and physicalism that account for the existence of consciousness (Chalmers, 1995, 1996). The "hard problem" remains to be solved, implying that the explanatory gap between materialism and qualia (Levine, 1983) remains.

The negative influences of this gap are profound. Other research involving consciousness is relegated to a situation wherein the research is either conducted with only a vague notion of what consciousness means or by neglecting consciousness outright. For instance, frustration with interdisciplinary integration has raised concerns about the future of cognitive science enterprises (Núñez, 2019). A fundamental error in cognitive science is that consideration of consciousness is neglected (Searle, 1990). Without consciousness, there is no cognition. In a sense, overall, the crux of all scientific research on consciousness points to the "hard problem" of consciousness. The problem comes full circle, and hence, must be faced directly.

This study endeavored to explore what consciousness is from the perspective of visual awareness, in a systematic manner[1]. First, a novel logic tool, called the non-identity law, was proposed based on physics to replace the law of identity. Second, the law of non-identity was applied to the analysis of expanded visual dynamics, building on a complete visual stimulation–response pathway related to the naturalistic observation of a night-shot still life. Finally, whether an out-of-body projection (OBP) is a physical behavior and the relations between an original, experience, and expression were discussed carefully. The results revealed that visual awareness, the representative sensation, can be understood in physical terms. Note that the term "visual awareness" in this paper is used in its broadest sense to refer to the entire process of sight, including various objective and subjective behavioral manifestations.

## 2 Theoretical preparation: the law of non-identity

Logic is known to be a cognitive tool for understanding the relationships among things in the universe, in which the law of identity, one of the three basic laws, is denoted as

$$Q = Q, \text{ or } Q \rightarrow Q,$$

where $Q$ is a thought object, such as a name, concept, event, and relation. The definition of the law is a tautology that states nothing at all about facts (Wittgenstein, 1921, pp. 63).

Here, we endeavor to challenge the law of identity within the frame of reference of natural science. Time, space, and matter are the three measurable interdependent elements that constitute the universe, while any object reveals itself (endurance, size, and position) by its interactions with other matter (Ridley, 1995, pp. 40–41, 85-86). Thus, an object can be denoted as

$$Q = Q(m, s, t),$$

where $Q$ is the name of the object. $Q$ depends on three measurable physical parameters, $s$, $t$, and $m$, in which $s$ denotes the shape and location of $Q$, $m$ the rest mass or relativistic mass (or the sum total of mass and energy) of $Q$, and $t$ the time. These values are determined by the interaction of $Q$ with its surroundings. For simplicity, unless otherwise stated, $Q$ refers to $Q(m, s, t)$ throughout the paper, and the following names, $Q'$ and $Q''$, are referred to similarly.

---

[1] If a proposition in a knowledge system can be neither proved nor disproved, the knowledge system is usually considered incomplete; accordingly, to solve the problem in a self-contained knowledge system, what we need to do is either complement (and or revise) the knowledge system or put the proposition in an expanded knowledge system (e.g. interdisciplinary integration) or both.

First, consider any two self-governing condensed objects, $Q(m, s, t)$ and $Q'(m', s', t)$, that are at rest and located in different places in an inertial frame of reference (Figure 1A). Macroscopic condensed matter consists of a large number of particles under four interaction forces. Suppose $Q$ contains $J(t)$ elementary particles $\{q_1, q_2, \cdots, q_{J(t)}\}$ at time $t$, where the $i$th particle $q_i$ possesses the mass $m(i)$ and the occupying space $s(i)$. Thus, $Q$ possesses the shape and location expressed by $s = \bigcup_{i=1}^{J(t)} s(i)$ and a mass denoted by $m = \sum_{i=1}^{J(t)} m(i)$. Similarly, $Q'$ containing $K(t)$ elementary particles $\{q'_1, q'_2, \cdots, q'_{K(t)}\}$ at time $t$ possesses the shape and location expressed by $s' = \bigcup_{i=1}^{K(t)} s'(i)$ and a mass denoted by $m' = \sum_{i=1}^{K(t)} m'(i)$. This is the basic physical representation of the two objects at the atomic level. Let us now return to the beginning of this paragraph and consider the problem: why is the number of objects in the inertial frame of reference deemed to be two?

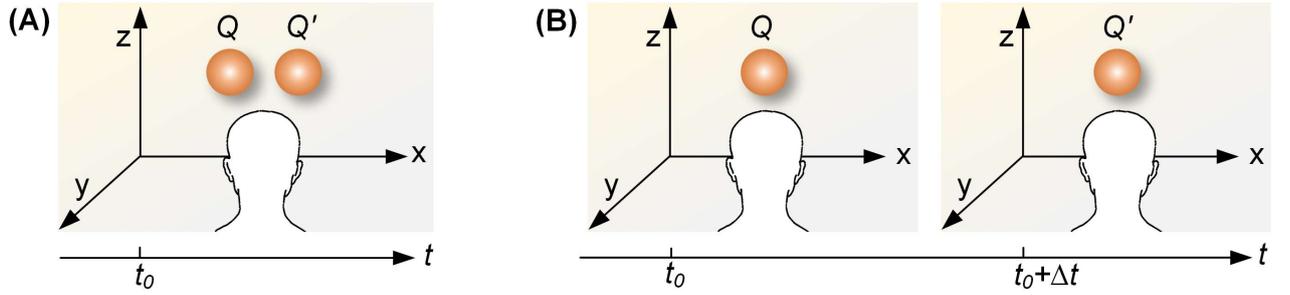

**Figure 1. Identifying an object in the inertial frame of reference. (A)** Discriminating two objects in the space dimension. An observer simultaneously observes two objects $Q$ and $Q'$ that are located at different places at the same time $t_0$. **(B)** Identifying one object in the time dimension. An observer sees an object $Q$ at $t_0$ for the first time; after closing his/her eyes for $\Delta t$, the observer sees an object $Q'$ at $t_0 + \Delta t$, where $Q'$ is the continuation of $Q$ in the time dimension.

Obviously, the spaces occupied separately by two macroscopic objects, whose borders and contours can be perceived by humans, have never overlapped with each other. In other words, the two macroscopic objects have no chance to meet or collide to exchange any substance with each other at any time; therefore, the two macroscopic objects do not share any particle at any time. Thus, the term "absolute non-identity law" can be defined as follows: two objects are called two objects (i.e., $Q \cap Q' = \phi$) if, and only if,

$$\forall (t_\xi) \big( s(t_\xi) \cap s'(t_\xi) = \phi \big),$$

where $Q$ and $Q'$ are the names of the two objects, respectively, $s$ and $s'$ are the spaces occupied separately by $Q$ and $Q'$, and $t_\xi$ is the time variable that traverses throughout the observation time, beginning with the moment $t_0$ and ending with $t_1$, i.e., $t_\xi \in [t_0, t_1]$. The absolute non-identity law shows how to distinguish one object from another. Note that although extracted from the inertial frame of reference with any two still objects, the absolute non-identity law can be generalized to the inertial frame of reference with any two objects in relative motion as long as their spaces are independent of each other at any one time.

Second, suppose that there is only one object in the inertial frame of reference (Figure 1B); the observer observes an object $Q(m, s, t)$ at $t_0$. Closing their eyes for $\Delta t$, the observer opens their eyes again and observes an object $Q'(m', s', t)$ at $t_1 = t_0 + \Delta t$ ($\Delta t \neq 0$). Is $Q'$ identical to $Q$?

Each spinning particle in $Q$ interacts with every other particle, resulting in the creation, decay, and annihilation of the particle whose position and momentum cannot be, even in principle, determined precisely (Ridley, 1995, pp. 112–113). Furthermore, substances have been continuously adsorbed onto $Q$ or have escaped from $Q$ in the open system. Considering that both the spatial structure and mass of $Q$ have been altered perpetually, the number of different particles in $Q'$ observed at $t_1$ is not identical to that of $Q$ at $t_0$. Thus, the term "relative non-identity law," can be defined as

$$\forall(t_0)\forall(t_1)(t_1 > t_0) \longrightarrow (Q' \cap Q \neq Q) \wedge (Q' \cap Q \neq Q') \wedge (Q' \cap Q \neq \emptyset),$$

where $Q'$ (i.e., $Q'(t_1)$) is the continuation of $Q$ (i.e., $Q(t_0)$) in the time dimension.

The absolute non-identity law provides a method for discriminating and naming one macroscopic object as distinguished from others in the perceivable space dimension[2], whereas the relative non-identity law indicates the evolution of one macroscopic object in the perceivable time dimension. Both laws, associated with the physical world, contain a greater reductionism connotation than the law of identity (in a sense, the law of non-identity is another expression reflecting the same aim as the law of identity). For example, by the relative non-identity law, we can readily solve some classic paradoxes, such as the ship of Theseus and the dispute about whether a man can step in the "same" river twice. However, since the absolute non-identity law, an unambiguous binary computing tool, is more rigorous than the relative non-identity law, we applied the former to the analysis of visual dynamics. In the following sections, we show that the non-identity law is essential as a reductionist tool for elucidating visual awareness.

## 3 Classic projection of brain-generated imagery from brain to observed object

### 3.1 Global visual dynamics model related to naturalistic observation

Visual awareness is a vital component of consciousness, and its biological substrate involves almost half of the cerebral cortex. It has several advantages over other components for investigating the neural basis of consciousness (Crick and Koch, 1990). In particular, it is scarcely affected by emotion. Significant progress has been made in elucidating visual dynamics in the brain (Walsh and Cowey, 1998; Lamme and Roelfsema, 2000; Tapia and Beck, 2014). However, the contribution of the messenger of the observed object and the outcome of the visual system have been neglected. Therefore, an analysis of the visual dynamics described with first- and third-person data (Chalmers, 2013) that arises over an expanded visual pathway, beginning with an observed object and ending with the visual outcome, is required. As this expanded visual pathway is global, it is also referred to as the "global visual pathway"; accordingly, its dynamics are called "global visual dynamics."

---

[2] The absolute non-identity law allows us to distinguish one thing from another, depending on their constancy and observability (appearances and bounds) in the space dimension, providing a bridge from the appearance of an object to the concept of the object. From the perspective of the knowledge system, the law of non-identity is the premise of all human cognition, including the law of excluded middle and the law of contradiction. If nothing is distinguishable in human perception, there is only one thing remaining, called "chaos."

Let us first build the global visual dynamics model. Suppose there is a ball in a dark room (Figure 2). The ball is denoted by $Q(m, s, t)$, whose definition is explained in the theoretical preparation section. At this moment, the ball is unknown to all observers; hence, it is referred to as the original or thing-in-itself. How does an observer perceive the original?

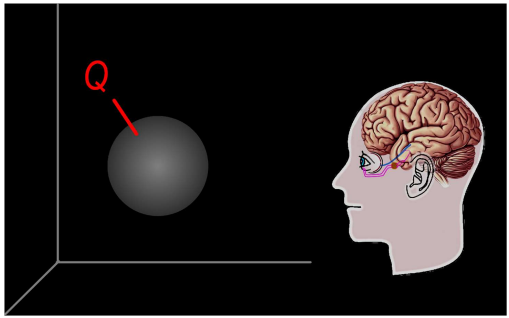

**Figure 2. Object in the dark, unknown to any observer**

It is well known that to explore an unknown thing actively, a stimulus–response test is usually required. Here, a specific research paradigm for the visual stimulus–response test was conceived, in which the single-pulse diffuse reflection is a dark-to-light-to-dark stimulus, while the human visual system acts as a detector. Different from self-luminous masking, not only is this paradigm closer to the naturalistic observation, but it also has an added advantage—the darkness does not enable the retina to encode signals effectively, thereby ensuring no interference before and after the visual stimulus. In compliance with the paradigm, two visual psychophysical experiments of perceiving night-shot still life under the pre-induced accommodation condition were conducted. Eighteen volunteers from the University of Electronic Science and Technology of China were recruited for the experiments (14 males and four females who were aged between 18 and 49 years, were not colorblind, and had no known neurological or visual disorders; six of them had normal vision and the others had corrected-to-normal vision. The study was reviewed and approved by the Ethics Committee of the University of Electronic Science and Technology of China. The participants provided their written informed consent to participate in this study before inclusion in the experiment). The experimental result shows that each participant correctly reported the direction, shape, and color of a constant or randomly selected still life after a single-pulse diffuse reflection stimulus lasting for 500 µs under the pre-induced accommodation condition associated with visual attention (see Supplementary Materials and Methods, Supplementary Text, Supplementary Figures S1–S4, and Supplementary Table S1 for details).

Having established a rough outline of the visual stimulus–response effect, let us now consider how the effect occurs in the detector (visual system). Based on the above visual psychophysical experiments, in conjunction with the neural dynamics in the extant literature, a complete visual dynamics model is presented (Figure 3). In this model, the flash, when turned on at $t_1$, begins to emit broad-spectrum photons that strike the surface of $Q$, where the photons are called detective photons. Some of the detective photons are absorbed and the rest are reflected around, some of which enter the eyes (Figure 3B). The process can be described as follows:

$$\xrightarrow{p_d(t_1)} Q(m, s, t) \xrightarrow{p_r(t_1)},$$

where $p_d$ represents the detective photons, and $p_r$ is the stray light cone entering the eyes, carrying the absorption and reflection information of $p_d$ interacting with $Q$ at $t_1$, i.e., the messenger light of $Q$. Given the complexity of particle physics, the space and mass parameters of the light were omitted for simplicity.

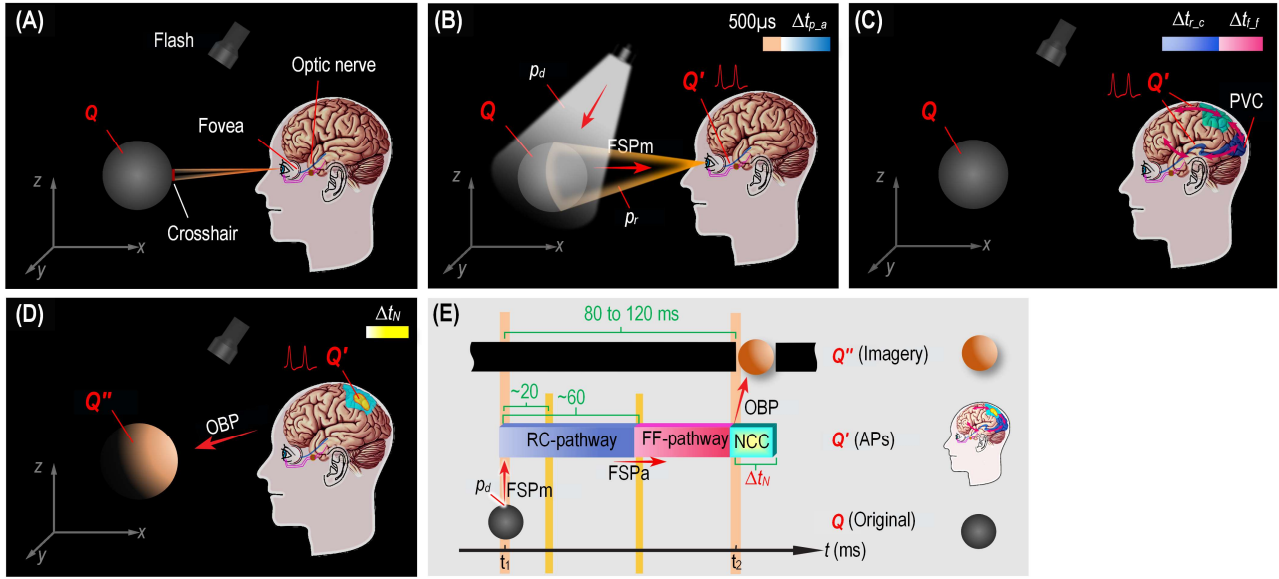

**Figure 3. Global visual dynamics in response to an afferent single-pulse messenger light signal (A)** An accommodation pattern is built in advance using a weak LED (light-emitting diode) crosshair, where the original $Q$ and human perception are all in the dark. **(B)** Messenger-light feedforward signaling pathway (FSPm). After the flash onset, some detective photons ($p_d$) reflected by $Q$, $p_r$ converging at the retina, have been encoded as action potentials (APs) ($Q'$) for $\Delta t_{p\_a}$. $Q'$ is submitted to the optic nerve. The perception is still dark. **(C)** AP feedforward signaling pathway (FSPa). $Q'$ has been relayed over the retinocortical pathway (RC-pathway) beginning at the retina and ending at the primary visual cortex in $\Delta t_{r\_c}$. Then, it takes $\Delta t_{f\_f}$ for $Q'$ to propagate sweepingly in a feedforward–feedback pathway (FF-pathway, red bidirectional long-dash ring). Finally, $Q'$ feeds back into the posterior parietal cortex (PPC, Cyan area), while everything is in the dark. **(D)** Out-of-body projection (OBP). After previous unconscious brain activity, a bright image $Q''$ arising from the PPC hot zone (the likely neural correlates of consciousness [NCC], yellow area) appears in front of the observer while the real world is still dark, and hence there is a recurrent OBP linking $Q'$ to $Q''$ via the NCC. Subsequently, everything falls into darkness. **(E)** Timing sequence of events related to vision. The reflected $p_d$ shaping FSPm at $t_1$ has been encoded as $Q'$ in the retina for about 20 ms (Kandel et al. 2012, pp. 594). Subsequently, $Q'$ has propagated through the rest of FSPa to the NCC. Eventually, the perception $Q''$ of $Q$ arises from the NCC at $t_2$. The duration of the unconscious response, $t_2 - t_1$, is between 80 ms and 120 ms, in which the retinocortical transmission time $\Delta t_{r\_c}$ is around 60 ms (Walsh and Cowey, 1998; Tapia and Beck, 2014).

Subsequently, the retina performs light-to-electricity transduction: the messenger light induces the rods and cones to trigger chemical signals continuously. These signals are sent through bipolar cells to the retinal ganglion cells where the chemical signals are eventually encoded as trains of action potentials (APs) that facilitate long-distance information propagation via the optic nerve (Kandel et al., 2012, pp. 577–601; Bear et al., 2015, pp. 293–330). Some ganglion cells show a response peak at only 20 ms after the flash onset (Kandel et al., 2012, pp. 594). The light-to-electricity transformation that takes place in the retina can be described as follows:

$$\xrightarrow{p_r(t_1)} retina \xrightarrow{Q'(e', s', t_1 + \Delta t_{p\_a})},$$

where $Q'$ denotes the encoding APs, $e'$ the bioelectrical energy, $s'$ the finite-dimensional traces of $Q'$,

and $\Delta t_{p\_a}$ the light-to-electricity transformation latency of about 20 ms[3]. Correspondingly, there is a messenger-light signaling pathway $Q \xrightarrow{p_r} retina$, an external feedforward signaling pathway (FSP), which is termed FSPm.

Up to this point, the retina has accomplished low-level visual processing tasks, such as light-to-electricity transduction, acquisition, and iconic memory. At this point, image acquisition devices can directly project an image onto a screen for us. In comparison, where and how does the human visual system present what one sees for oneself?

First, the encoding $Q'$ arising from the retina is conveyed through the lateral geniculate nucleus to the primary visual cortex (Figure 3C), whose neural pathway is often referred to as retinofugal projection (Bear et al., 2015, pp. 333). Electrophysiological research has demonstrated that the retinocortical transmission time $\Delta t_{r\_c}$ varies from 55 ms to 70 ms and averages approximately 60 ms (Tapia and Beck, 2014).

Next, from the primary visual cortex, the encoded information is transmitted via two major pathways, first described in a monkey: a dorsal pathway into the frontal lobe through a number of extrastriate areas in the parietal lobe, and a ventral pathway into the frontal lobe through the inferior temporal cortex (Mishkin et al., 1983). Generally, the dorsal pathway processes the spatial information (position, motion, speed) related to fast visuo-motor control, whereas the ventral pathway processes information about the form (color, shape, texture) related to perception, both of which also exist in other sensory systems (Kandel et al., 2012). Additionally, the propagation of APs through lateral and feedback connections, in which the same cortical neurons contribute to different analyses at different moments in time, can be incorporated into a sweeping feedforward–feedback response (Lamme and Roelfsema, 2000).

Note that although unconscious, the above stage where $Q'$ propagates sweepingly over the intracorporal FSP with conjunction and disjunction, is not only correlated with consciousness, but is also necessary for it (Tapia and Beck, 2014). Compared to image acquisition devices, the added sweeping feedforward–feedback response demonstrates that the brain makes an extraordinary effort to generate imagery.

At length, the distributed encodings linked to the different visual features (e.g., position, shape, and color) are eventually bound together (Crick and Koch, 1990, 2003; Treisman, 1996). As a result, a perception of $Q$ with its various features arises from the NCC. The NCC is believed to be primarily localized to a posterior cortical hot zone that includes sensory areas (Koch et al., 2016); however, the distribution of the NCC remains controversial (Mashour, 2018). Considering that no consequence is altered by variations in the layout of the NCC, the statement of the posterior cortical hot zone was used in this study.

Thus far, the visual signaling pathway, $p_d \rightarrow Q \xrightarrow{p_r} retina \xrightarrow{Q'} NCC$, has been completely presented. In contrast to the extraneous messenger-light pathway, the AP pathway $retina \xrightarrow{Q'} NCC$, the intracorporal part of the FSP is termed FSPa. Previous research investigating whether perception is blocked by transcranial magnetic stimulation pulses applied to the early visual cortex demonstrates that the duration of unconscious brain activity after stimulus onset, $\Delta t_{FSPa}$ (i.e., $\Delta t_{r_c} + \Delta t_{f\_f}$), typically ranges from 80 ms to 120 ms (Walsh and Cowey, 1998; Tapia and Beck, 2014), where $\Delta t_{f\_f}$ is the duration of the feedforward–feedback response. This observation is consistent with findings from other visual psychophysical studies. For instance, the critical flicker fusion frequency of humans is 60 Hz (Healy et al., 2013), while perception occurs approximately 80 ms after stimulus onset in the flash-lag effect experiment (Eagleman and Sejnowski, 2000). In other words, these experimental methods for measuring the cycle of visual dynamics are equivalent.

---

[3] Any substance takes time to move from one position to another. Its speed cannot exceed the speed of light ($c_0$=299792458 m/s). Here, the travel of light also requires time, but due to the short transmission distance, the transmission time of light is negligible. In comparison, the duration of light-to-electricity transformation and the period of action potentials traveling from one site to another are significant.

The visual experience of an afterimage can persist for several minutes, helping us to infer the duration of the NCC activity without examining a biomarker of NCC activity. Contrastingly, the duration of visual perception of a night-shot still life image is very short, and therefore the duration of NCC activity, $\Delta t_N$, should be accurately determined by checking the NCC activity. However, the NCC remains to be finalized; hence, $\Delta t_N$ remains unknown. Instead, another issue of greater concern is the output of physical visual processing.

It should be considered that the migrating APs carrying all visual information on $Q$ are situated inside the brain (Figure 3C); hence, the output of visual processing should likewise be confined to the brain. However, interestingly, although the real world in this experiment is dark, an orange ball $Q''$ appears in front of the observer, rather than inside the brain of the observer (Supplementary Figures S3–S4, and Figure 3D). Although short-lived, $Q''$ is a real visual experience for the observer. No matter what it is, $Q''$ can always be expressed as

$$Q''(m'', s'', c, t_2),$$

where $Q''$ denotes what the observer sees in front of him/her; it can be described by four parameters: $m''$, the unknown mass of $Q''$; $s''$, the space occupied by $Q''$; $c$, the color of $Q''$; and $t_2$ (i.e., $t_1 + \Delta t_{FSPa}$), the time at which the perception emerges from the NCC (Figure 3E).

There is no doubt that a transformation (also called emergence) has occurred in the NCC:

$$\xrightarrow{Q'(e', s', t_2)} NCC \Rightarrow Q''(m'', \ s'', c, \ t_2).$$

Analogous to the projectors, the behavioral manifestation, $NCC \Rightarrow Q''$, can be termed an OBP. So far, we have mainly presented the global visual dynamics, based on which we can conclude that the visual system possesses an FSP-OBP pathway,

$$p_d \to Q \xrightarrow{p_r} retina \xrightarrow{Q'} NCC \Rightarrow Q'',$$

where $p_d \to Q \xrightarrow{p_r} retina \xrightarrow{Q'} NCC$ is the unconscious FSP, $NCC \Rightarrow Q''$ is the OBP pathway of the conscious outcome, and the OBP is the brain-generated imagery in response to the afferent messenger light of a particular original.

### 3.2 Visual dynamics analysis with non-identity law

Let us now analyze the relationships between original $Q$, encoding $Q'$, and OBP $Q''$. First, the encoding $Q'$ of the messenger light and the original $Q$ are located at both ends of the FSPm pathway, i.e., $Q$ at rest is located at an extracorporeal place, while $Q'$, in response to some detective photons reflected by $Q$, is located inside the body. Hence,

$$\forall(t_\xi)\big(s(t_\xi) \cap s'(t_\xi) = \emptyset\big),$$

where $s$ and $s'$ are the spaces owned separately by $Q$ and $Q'$, and $t_\xi$ is the time variable that traverses throughout the observation period. By the non-identity law described in the theoretical preparation section, we obtain

$$Q \cap Q' = \emptyset,$$

i.e., $Q'$ and $Q$ are two different things.
Similarly,

$$Q'' \cap Q' = \emptyset.$$

$Q''$ and $Q'$ are, therefore, two different things too. Then, what is $Q''$? This could simply be answered by announcing a new concept. However, after searching the existing literature, we are convinced that an OBP, the conscious outcome of the visual system, i.e., the brain-generated imagery, is the so-called visual awareness in a narrow sense. In other words, what we see in front of us is precisely our visual awareness. Notably, although we know the timing sequences of $Q$, $Q'$, and $Q''$, the spatial connection between $Q$ and $Q'$, and the spatial connection between $Q'$ and $Q''$, we are unable to deduce the spatial connection between $Q''$ and $Q$ directly by the equations $Q \cap Q' = \emptyset$ and $Q'' \cap Q' = \emptyset$. Consequently, the next step is to confirm their spatial connection quantitatively.

### 3.3 Projection position reciprocally determined by vision and touch

As described previously, the dorsal pathway determines where the external thing is, whereas the ventral pathway determines what it is. Visuospatial information regarding gaze direction and gaze distance is common to all visual cortical areas (Dobbins et al., 1998). The messenger light of an original, however, cannot carry visuospatial information other than the absorption and reflection information of detective photons interacting with an original. Thus, it is natural to question where the visuospatial information originates.

The visual system possesses a mechanism to generate the visuospatial information by itself, involving a set of sophisticated dynamic optical control mechanisms comprising eye movement, pupillary reflex, and accommodation (Kandel et al. 2012, pp. 562). In principle, a dynamic system can be regarded as an automatic regulatory system. When the mechanism is introduced into the global visual dynamics model, a symbolized semi-closed-loop visual dynamics model, the automatic regulatory system with two typical transformations that occur separately at the eye and NCC can be presented (Figure 4A). In this symbolized model, an optical focus is achieved by applying the top-down encoding $Q'_{mov}$ for eye movement and the local encoding $Q'_{acc}$ for accommodation to the extraocular and ocular muscles, where the encodings are correlated with the direction and distance of the observed object $Q$. In turn, the visuospatial information on the direction and distance of $Q$ can be represented by both encodings. Thus, it can be inferred that the sites at which $Q'_{mov}$ and $Q'_{acc}$ are encoded for an optical focus provide a copy of the encodings to all visual cortical areas for intermediate- and high-level visuospatial processing, in compliance with size constancy in terms of shape, motion, and speed (Kandel et al., 2012, pp. 602–637; Schwartz et al., 1983).

In fact, spatial cognition is achieved by visuotactile integration training (Batista et al., 1999; Gentile et al., 2011; Chen et al., 2016). When touching an object, one can shape a tactile experience $Q''_{tou}(p''_{tou}, s''_{tou}, k''_{tou}, t)$ of $Q$, where $p''_{tou}$ is the perceptual pressure in response to the electromagnetic interaction that occurs at the contact interface and $k''_{tou}$ is the perceptual temperature of $Q$. Although many more objective and subjective parameters can be adopted, only four parameters are used here for simplicity. Obviously, the tactile outcome is not located inside the brain, but at the contact interface. Therefore, similar to the symbolized visual dynamics model, we can obtain a symbolized closed-loop tactile dynamics model: the automatic regulatory system with two typical transformations that occur separately at the NCC and the finger receptors (Figure 4B).

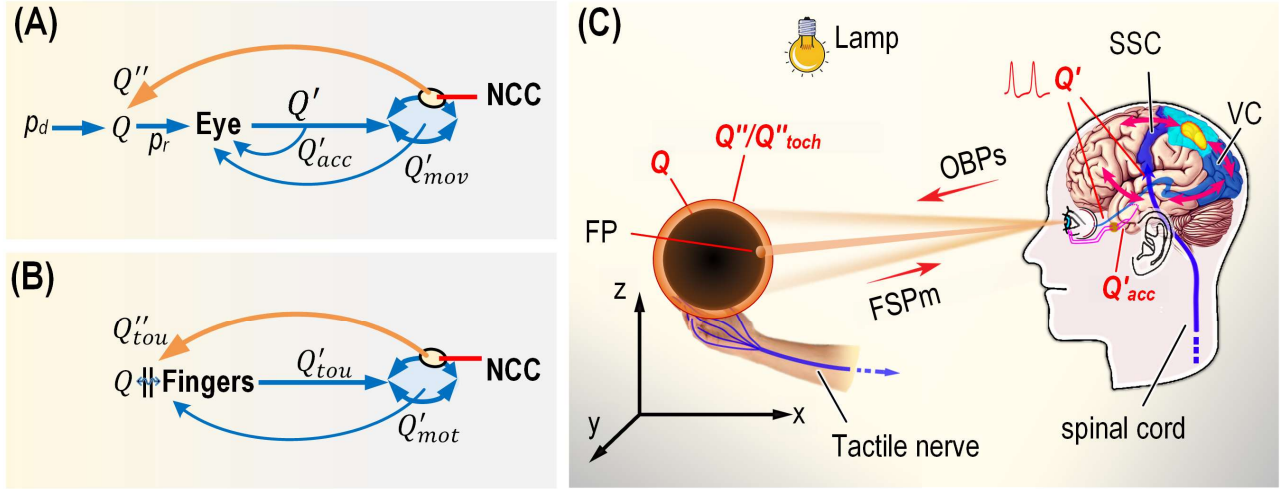

**Figure 4. Visual-tactile intercalibration based on visual and tactile automatic regulatory systems (A)** Semi-closed-loop visual regulatory system. When an observer stares at $Q$, the encoding $Q'$ of the messenger light of $Q$, which arises from the retina, is sweepingly propagated to the feedforward–feedback circuitry (ellipsoid). Consequently, a top-down encoding $Q'_{mov}$ for eye movement control innervates the extraocular muscles to move the eyeball, pointing the fovea towards the fixation point (FP) on $Q$, whereas the local encoding $Q'_{acc}$ for accommodation, which derives from the optic nerve, innervates the ocular muscles to regulate the pupil and lens. As a result, the controlled optical system only allows a narrow beam of messenger light of the FP to focus on the fovea and collaterally allows the messenger light of $Q$ to focus on the retina. At length, a sharp image $Q''$ is projected back to $Q$ by the neural correlates of consciousness (NCC). **(B)** Closed-loop tactile regulatory system. The encoding $Q'_{mot}$ coming from the motor area controls the coordinated arm–wrist–finger motion until the fingers touch an object. Consequently, the encoding $Q'_{tou}$ of the electromagnetic interaction occurring at the interface $Q \parallel$ Fingers is fed back through the feedforward–feedback circuitry into the NCC, by which a tactile experience $Q''_{tou}$ is projected to the interface. **(C)** Visual–tactile intercalibration. The reflected photons shaping the messenger-light feedforward signaling pathway (FSPm) reach the retinae, at which $Q'$ is encoded and conveyed to the visual cortex (VC). The tactile signal encoding, through the tactile nerve and spinal cord, enters the somatosensory cortex (SSC). Both signals propagate sweepingly through the feedforward–feedback pathway to the NCC. At length, an image $Q''$ is projected onto $Q$ and is confirmed by the tactile perception $Q''_{tou}$.

In contrast to the semi-closed-loop visual system, tactile receptors, the farthest node of the closed-loop pathway of touch, can directly approach an object through the moveable arm and fingers (Figure 4C). For example, to explore the outside world, the top-down encoding $Q'_{mot}$, encoded by the motor system, voluntarily drives the arm and fingers until the fingers touch an object. Hence, there is a contact interface $Q \parallel$ Finger at which the electromagnetic interaction occurs. Subsequently, the encoding $Q'_{tou}$ of electromagnetic interaction feeds back into the brain and mediates the NCC to project a tactile perception $Q''_{tou}$ to the contact interface. Thus, $Q''_{tou}$ spatially accords with $Q$, i.e., $s''_{tou} \approx s$. Moreover, there is evidence that spatial information of the same object obtained by sight and touch can be calibrated with each other (Chen et al., 2016), implying that $s'' \approx s''_{tou}$. Therefore, $s'' \approx s$. This demonstrates that the visual OBP is spatially superimposed onto the observed object, creating a complete visual causal chain as shown in Figure 4A.

## 4 Nontrivial projection of brain-generated imagery separated from its original

As described previously, a classic visual pathway, $p_d \to Q \overset{p_r}{\to} retina \overset{Q'}{\to} NCC \Rightarrow Q''$, where $Q''$ is usually superimposed onto its original $Q$ can be reciprocally verified by visual and tactile perception, facilitating the belief that what one sees is exactly the object itself. However, another instance that we typically encounter on a daily basis should challenge this impression: the mirror image.

To revisit the mirror image, let us conceive a thought experiment on the mirror image (Figure 5A). When an observer looks into a mirror, the messenger-light pathway of $Q$ is $p_d \to Q \overset{p_r}{\to} mirror \overset{p_r}{\to} retina$, where the FSPm is deflected by the mirror. Hence, there are two segments of the messenger-light pathway: $S_1$, a pathway from $Q$ to the mirror, and $S_2$ from the mirror to the retina. The messenger light converging on the retina is encoded as $Q'$, which enters the brain through the optic nerve for intermediate- and high-level visual processing. To present the image $Q''$ behind the mirror clearly, the top-down encoding $Q'_{mov}$ innervates the extraocular muscles to point the fovea to $S_2$, and the local encoding $Q'_{acc}$ innervates the ocular muscles for accommodation, so that the optical system focuses the messenger light of the fixation point on the fovea. Eventually, the observer sees a sharp image $Q''$ behind the mirror.

Let us now withdraw the mirror and box and move the ball to the place where the preceding $Q''$ used to be, while leaving the rest of the scenario unchanged (Figure 5B). When the observer fixates on the current fixation point (i.e., the previous virtual fixation point), the terminal pathway of the current

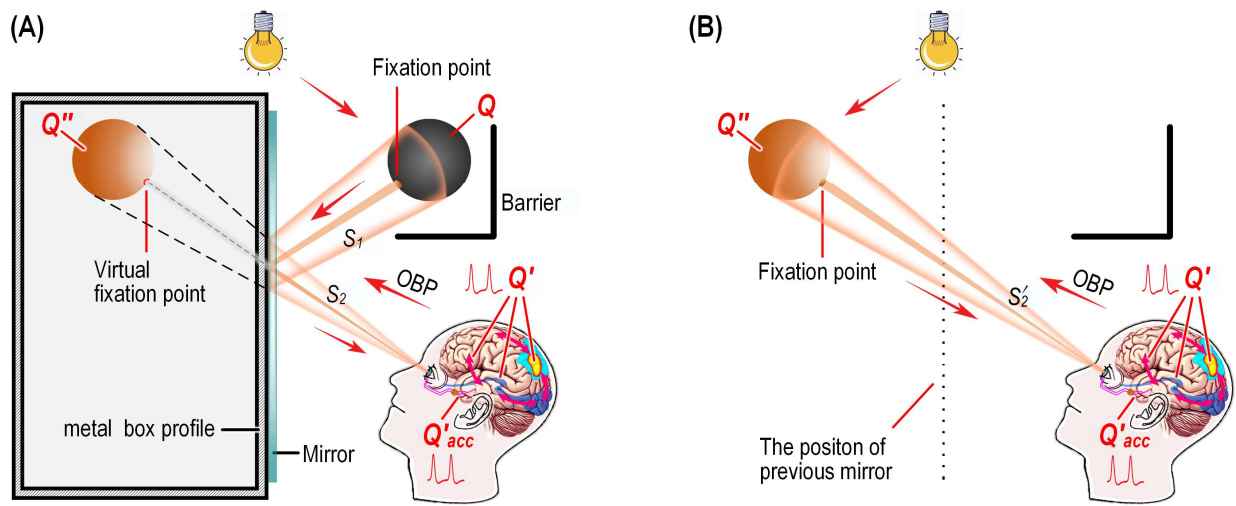

**Figure 5. Separating visual projection from its original by regulating the afferent messenger-light pathway with a mirror (A)** A mirror is fixed on one side of a sealed metal box, whereas a ball $Q$, covered by a semi-enclosed barrier hangs above the observer. The lamp, when turned on, fires detective photons whose messenger-light feedforward signaling pathway (FSPm) from the unseen object $Q$ to the retina of the observer is divided into two segments, $S_1$ and $S_2$. The messenger light in $S_2$ induces the retina to fire action potentials (APs) ($Q'$). $Q'$ enters the AP feedforward signaling pathway (FSPa), from which the top-down encodings $Q'_{mov}$ (not shown, see Figure 4A for detail) and the local encoding $Q'_{acc}$ arise to regulate the optical system (Kandel et al. 2012). As a result, the messenger light of the fixation point (the dark brown centerline of $S_1$ and $S_2$) is focused on the fovea, and the messenger light of $Q$ is focused collaterally on the retina. Eventually, a sharp reflection $Q''$ of $Q$ appears behind the mirror. **(B)** After the metal box and mirror are withdrawn, the observer watches the ball again (here, the original $Q$ is not shown; see Figure 4C for detail).

FSPm, $S_2'$ is in line with the previous $S_2$; accordingly, the top-down encoding $Q_{mov}'$ and local encoding $Q_{acc}'$ for the current FSPm are the same as those for $S_2$. The same is true for the intermediate- and high-level visuospatial processing. Consequently, in either case, the NCC projects the same image to the same place; i.e., both observation modes are equivalent for visual perception. To distinguish the two types of OBP, the OBP that is spatially superimposed onto its original is termed classic projection, whereas other OBPs, such as starlight deflection, mirror image, afterimage, imagination, and dreams, are termed nontrivial projections.

More importantly, in contrast to the classic projection, the mirror image clearly shows that

$$\forall (t_\xi)(s(t_\xi) \cap s''(t_\xi) = \emptyset),$$

where $s$ and $s''$ are the spaces owned separately by $Q$ and $Q''$, and $t_\xi$ is the time variable that traverses throughout the observation period. Based on the non-identity law, we obtain $Q'' \cap Q = \emptyset$, revealing that both still objects, the visual projection $Q''$ and the original $Q$, are two different things. It is noteworthy that $Q''$ and $Q$ are deemed to be symmetric with respect to the mirror; however, this cannot be experimentally determined. For this reason, a thought experiment is adopted instead of a practical experiment.

Moreover, the mirror image provides evidence that the OBP pathway, in response to the messenger light of the original, has constancy, and that the OBP can be manipulated by regulating or reconstructing the FSPm. The FSPm-regulating technique, whereby FSPm is regulated in real time has long been used to create optical tools, such as the microscopes, telescopes, periscopes, and spectacles, whereas the FSPm-reconstructing technique, whereby FSPm can be reconstructed anytime and anywhere, is used to make image and video productions, such as 3-dimensional paintings and movies, and virtual reality. Additionally, the FSPm-regulating technique can be used to treat chronic neurological disorders, such as phantom limb pain and hemiparesis caused by stroke (Ramachandran and Altschuler, 2009). It can even be used to unveil a nontrivial mental phenomenon, the out-of-body experience (Ehrsson, 2007; Lenggenhager et al., 2007). In summary, these seemingly isolated manifestations are just different forms of nontrivial projections whose FSPm are regulated, reconstructed, or simulated.

To understand the classic and nontrivial projections further, we constructed a scene where one ball and two mirror images appeared simultaneously (Figure 6), prompting the question: which one is true? Based on the OBP principle, the ball that could be touched was considered a classic projection, whereas the two other mirror images were nontrivial projections; therefore, all of them were true images but none of them was the real object[4]. Based on the non-identity law, the three images located at different positions were three things. There is, however, little doubt that all of them, which were very much alike in appearance, corresponded to the same unknown original. This demonstrates that an original can simultaneously present multiple images to an observer.

---

[4] All the images are regarded as "true," because perception is the starting point of human cognition; i.e., man is the yardstick of everything. In other words, what one perceives, even a mirage, is a "true" perception. Notably, "true" image does not mean that its original actually exists, e.g., computer-generated imagery.

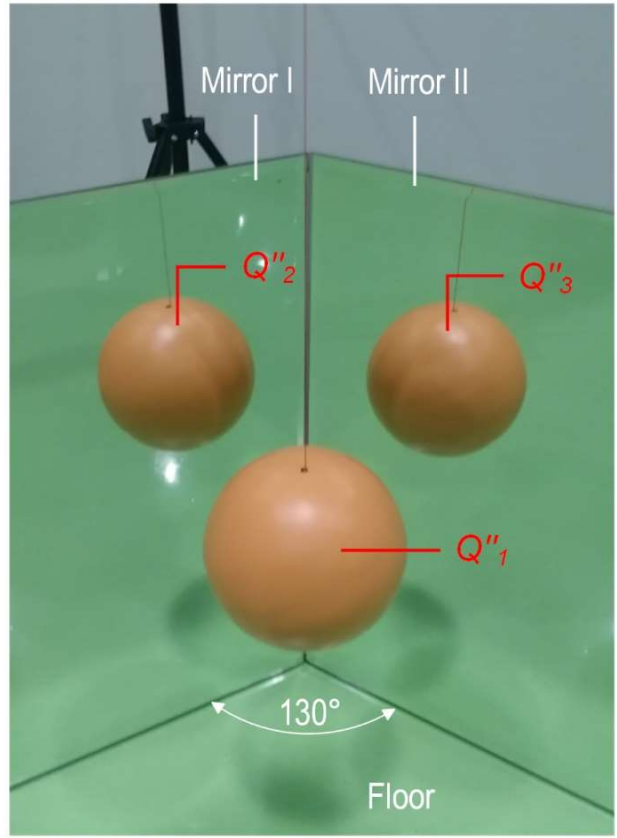

**Figure 6. Multi-image perception of an original.** The test equipment comprised two mirrors (56 × 35 cm, L×W) and an orange metal ball with a diameter of 15 cm. The mirrors are close to each other and have an intersection angle of approximately 130 degrees. The ball is suspended approximately 25 cm above the floor. $Q_1''$ is a "ball" that can be seen and touched simultaneously, whereas $Q_2''$ and $Q_3''$, which cannot be touched, are two nontrivial images whose messenger-light feedforward signaling pathways are regulated by mirror I and mirror II, respectively. The three images possess the same shape and color but differ in their positions and directions. Note that $Q_1''$, which could be touched by an observer, is generally regarded as a real object; however, it is merely a classic projection of brain-generated imagery in response to the messenger light of an unknowable original.

## 5 Discussion

### 5.1 Is visual projection a physical behavior?

Scholars have long been aware of the OBP of an afterimage (the afterimage can be projected anywhere and the size increases with the projection distance), and have proposed Emmert's law (Boring, 1940), which provides a nontrivial cue for revealing visual awareness. Perhaps, because subsequent research has focused on the feedforward size constancy instead of visual projection (Epstein et al., 1961), and there is an unexplainable deviation occurring between the hypothesis and fact (Young, 1948; Furedy and Stanley, 1970), the visual projection has been regarded as an isolated visual phenomenon and has not been developed into a universal visual principle, until now. The root cause, however, may lie within consciousness itself.

Consciousness, generally considered the subjective experience arising from the brain, has several typical components, such as sensation, emotion, reasoning, imagination, and self-awareness, which possess some common features, such as subjectivity and intentionality (Searle, 2004, pp. 133–145, 1984, pp. 13–27). Then, what about visual awareness as a representative sensation?

First, visual awareness is based on an FSP-OBP pathway whose FSP has a physical delay of approximately 100 ms after flash onset. Similarly, related research on motion control has demonstrated that a decision can be encoded into the prefrontal and parietal cortex up to 10 seconds before it enters awareness (Soon et al., 2008). Furthermore, a lesion to the FSPa may pose a loss of conscious outcome, such as blindsight and motion blindness (Lamme, 2001; Goodale and Milner, 2013; Zihl et al., 1983), resulting in a remarkable deterioration in behavioral flexibility. However, even though the FSP functions normally, all brain activities do not cause conscious awareness (Custers and Aarts, 2010). These provide converging evidence that the postponed conscious outcomes of sight and motion are determined by sophisticated brain activity, supporting the argument that conscious outcome is neither predictive nor online, but rather postdictive (Eagleman and Sejnowski, 2000). It also supports the hypothesis that conscious outcome is a collateral product or epiphenomenon of brain activity (Huxley, 1874, pp. 240), similar to a host-controlled screen that selectively displays something. Although this view is not widely accepted, this principle is used to conduct neurological investigations into and clinical treatments for patients with cognitive disorders (e.g., Alzheimer's and Parkinson's disease) and to anesthetize a patient before an operation to deprive the patient of pain and even awareness.

Second, the media of the FSP ($p_d \rightarrow Q \xrightarrow{p_r} Retina \xrightarrow{Q'} NCC$) are known to be photons and APs in physical terms. Analogously, we assume that the medium of the OBP ($NCC \Rightarrow Q''$) is $x$; hence, the updated OBP expression is $NCC \xRightarrow{x} Q''$. This prompts the following question: what is $x$? The materials that support life are the same as those present in the universe. Since the operation of an object can be influenced by its interaction with other objects (Ridley 1995, pp. 69–86), we can always use certain materials to intervene with the OBP if the visual projection is a physical behavior, as we have done in other scientific experiments. However, we have never encountered a situation where a wall (or a sealed metal box, or anything else) on which a mirror is fixed blocks the OBP pathway and obstructs the visual perception of the mirror image. Moreover, the color of the object, whether it is inside or outside the mirror, cannot be deduced from the three fundamental concepts in physics. Therefore, a mirror image is not a physical thing and OBP is not a physical behavior, thereby falling into the so-called phenomenological or transcendental category (Kant, 1783). It is worth emphasizing that the method can be called a mirror test, which is likewise suitable for examining theories of consciousness.

Third, it should be highlighted that the brain-generated imagery is projected back onto an original in response to the afferent messengers of the original, signifying that the NCC can perform voluntary OBP, whose behavior can be termed projection intentionality. This evidence indicates that humans have an instinct for producing subjective OBP to keep up with its original in space, like a searchlight trying to track and shine on a moving object in the dark, or a bat positioning and catching a moth in the dark. The same is true for other sensations, such as touch and hearing; for example, when we touch something with a tool, we can feel the touch at the tip of the tool, as an extension of the body (Miller et al., 2018; Heed, 2019), signifying that touch also possesses the capacity for OBP, extending beyond the boundary of the body. Accumulating evidence has demonstrated that human sensations, such as vision, touch, and hearing, possess the attribute of spatial expansion outward from the body. Thus, it can be inferred that intentionality is one aspect of the subjectivity of consciousness and that the recurrent OBP may be the psychophysical origin of the intentionality of consciousness.

Taken together, all experiences, the responses to the messengers of the originals or to their memory recall, should fall into the subjective manifestation category. For example, "space," "time," and "acting force," generally regarded as so-called objective things, are merely the measurable components of conscious outcomes that can be reciprocally verified by different perceptions (particularly, visual and tactile perceptions), suggesting that what one perceives is simply a fantasy that coincides with reality; i.e., none of the experiences are original but are merely subjective perceptions that metaphysically reflect particular originals via the afferent messengers of the originals[5]. Hence, it is not difficult to understand why the symmetry of the mirror image has never been experimentally determined and why the visual projection presented in afterimages has not been developed into a universal visual principle.

## 5.2 Isomorphic relations between original, experience, and expression

Thus far, a panorama of the universe has unfolded: the world for an observer (first-person) is composed of dark originals, the originals' messengers that can be perceived by the observer, the body in which two transformations occur, and a subjective perception (conscious outcome, also known as experience, phenomenon, imagery, appearance, manifestation, or representation) that follows or reflects the dark originals. This corroborates the following well-known philosophical ideas: "although the thing is completely unknown to us as to what they may be in themselves, we know through the representations which their influence on our sensibility, and to which we give the name of a body" (Kant, 1783, pp. 40); "manifestation stems from thing-in-itself. … thing-in-itself and its manifestation are the same, though they are named differently, which are called mysteries, and the mystery underlying the mysteries is the gateway to all understanding" (Lao-Tzu, 480 BC). Lao-Tzu further argued that constant manifestations can be used to investigate the bounds of a thing, which coincides with the basis of the non-identity law. Interestingly, this argument is supported by the discovery that neurons performing visuospatial processing at different levels are most sensitive to the borders and contours of the observed objects in a scene (Kandel et al., 2012, pp. 556–653).

However, not all messengers can induce sensory receptors to shape an experience. For example, invisible originals, such as air and most types of radiation, provide evidence that in some cases, human sensory experience is absent from the operation of varied originals[6]. Additionally, the nontrivial projection mentioned above demonstrates that the experience may deviate markedly from its original. Therefore, human experience is a limited and even nontrivial reflection of the originals in some cases. It must be emphasized that a sensory experience, i.e., the response to the messengers of a particular original, is trustworthy, yet also questionable.

Since an experience, which arises from the NCC and fades as the messengers disappear, is a transient, private, and subjective event, how do we record and express our feelings and communicate with each other? We have to use certain languages (gestures, sounds, characters, symbols, or drawings) that can be perceived by others using sight, hearing, and touch, to describe, name, and record the phenomenological events, and statistically develop a "constant conjunction" between different events (Hume, 1784, pp. 53–57). This process is labeled phenomenological expression, which is a man-made, abstract, and limited mapping (isomorphism) of the experience. As such, it is replete with symbolization and suggestiveness. Interestingly, in turn, the symbolic and kinematic suggestiveness of the expression leaves something to our imagination, resulting in several rigorous metaphysical systems, such as mathematics, logic, and graphic art, which have laid the groundwork for scientific research.

---

[5] Time, space, and matter are not only scientific issues but also classic philosophical issues. These issues concern the nature of the universe and its motion. "The flying arrow is at rest," a well-known paradox proposed by Zeno (Huggett, 2019), is yet to be resolved systematically and quantitatively.

[6] Each species has its own limited but sufficient sentience for its survival, which is the inevitable result of evolution. The same is true of the selective perception of humankind. Too much sentience would otherwise only bring about confusion and waste. Even if one species historically had other forms of sentience, lack of use usually resulted in the degradation of that sentience.

We can go even further; using the above metaphysical systems, we can consolidate as many manifestations within a unified metaphysical framework as possible. This process may be termed metaphysics-consolidated expressions, such as conservation laws, evolution theory, Maxwell's equations, special relativity, and the standard model. In parallel, through the phenomenological observations by the aid of advancements in experimental apparatus combined with proper rational analysis, scientists have made many significant empirical findings such as elementary elements, elementary particles, the constant speed of light, and the DNA double-helix structure.

In comparison with the phenomenological and metaphysics-consolidated expressions, it is more revolutionary to seek the cause of the manifestations. First, regarding certain manifestations as evidence, we tentatively put forward a hypothesis as the common cause for the manifestations and model it with several acknowledged constant laws, such as interaction laws, conservation laws, and evolution rules. Next, if the deduction or laboratory development on the hypothesis does not fit into the current and hypothetical-deductive-forecasted evidence, we revise the hypothesis until it fits the evidence well. Subsequently, the best-matching hypothesis is regarded to be true. This process is labeled hypothetical expression. Interestingly, it is the hypothesis-to-manifestation research approach that has helped us break through cognitive limitations and that fuels the progress of science, resulting in a series of theories, such as heliocentrism, gravity, atomic model, energy quantum, and general relativity.

However, both phenomenological and hypothetical expressions remain tentative. A manifestation is only a limited or even a nontrivial reflection of the real world; therefore, the phenomenological expressions extracted from the manifestations are fallible, which further affects the legitimacy of metaphysics-consolidated expressions. The hypothesis-to-manifestation expressive paradigm may put our cognition or metaphysics at risk of straying from the natural original-to-manifestation route. It is, therefore, inevitable and unsurprising that multiple hypotheses related to the same issue may exist in parallel for a long time. Unfortunately, we have no other choice. These provide insight into the core thought of Taoism: "we can name and describe manifestations, think and talk about Tao underlying manifestations, but the expressions are not manifestations and Tao themselves and, consequently, are fallible; ... nevertheless, we can still explore the mystery of Tao and the bounds of manifestations by their constancy" (Lao-Tzu, 480 BC).

Since manifestations and expressions are fallible, how do we ensure that our intellectual adventure is safe? In addition to rigorous algorithmic rules, the expressed hypotheses or truths should satisfy the criteria of compatibility, completeness, and simplicity. As far as the mind–body problem is concerned, Kant's transcendental philosophy and Lao-Tzu's Wu-You thought, which acknowledge the existence of unknowable things in themselves, are not only compatible with each other but are also applicable to classic and nontrivial imaging. They are thoroughly materialistic philosophies[7] (in fact, most of us usually see the things before our eyes as real things and, hence, are idealists). Contrastingly, parallel traditional philosophies, such as dualism, mechanical materialism, and idealism, which treat free will as the cause of human activities to varying degrees, are neither compatible with each other nor able to pass the mirror test (that is to say, none of these traditional philosophies are able to explain the phenomena shown in Figure 5A and Figure 6). Fortunately, subjective perception generally conforms to its original. However, this is not always the case. Therefore, we should be open to the expressions with respect to the manifestations and their underlying causes.

---

[7] Things in themselves refer to the dark unknowable things in the objective originals' world that underlies the subjective phenomenological world, including the dark things underlying the perceptions of matter, space, and time. Literally, therefore, the word "materialistic" is an incomplete makeshift expression and should be completed in the future.

# 6 Conclusion

The current study primarily sought to address what consciousness consists of from the angle of visual awareness. It consists of a comprehensive investigation of the global visual dynamics between the thing-in-itself, brain, and visual experience, via an interdisciplinary approach involving philosophy, physics, logic, neuroscience, and psychophysics. The results revealed that visual awareness involves a postponed, recurrent OBP, suggesting that the visual system has an instinct of not only subjectively imaging, but also projecting the image back onto its original, according to the cue of the afferent messenger-light pathway of the original. This finding, coupled with the NCC, adds a key puzzle piece to the visual system, forming a complete visual causal chain. Contrastingly, a lack of OBP may result in a remarkable deterioration in behavioral flexibility, such as blindsight and motion blindness. A possible explanation for this may be that the OBP is an optimum evolutionary strategy and is essential for the survival of fast-moving creatures.

In light of this finding, this study clearly explained nontrivial projections such as the mirror image, suggested the mirror test as a criterion for examining theories of consciousness, elucidated the psychophysical root of both subjectivity and intentionality of consciousness, and highlighted a growing principled understanding of the isomorphic relations between the original, experience, and expression. These results suggest that under adopted consciousness, the interdisciplinary integration of cognitive science can be further promoted and that the debate regarding the bounds of artificial intelligence, i.e., whether a machine can have consciousness, may be settled. Although OBP plays a crucial role in visual awareness, the recurrent OBP pathway is only roughly provided in this study, while the only related hypothesis, Emmert's law, is at variance with the fact. Therefore, the projection geometry of the OBP remains to be experimentally determined. The solution to the projection geometry will further prove that consciousness, including self-awareness, can be understood in physical terms. Further research in this field will require the interpretation of another aspect of the "hard problem" — how consciousness arises from the brain activity of life developed from a fertilized egg, and empirical studies that address issues regarding more accurate dynamics of sensing, emotion, and thinking.

# 7 Conflict of Interest

The authors declare that the research was conducted in the absence of any commercial or financial relationships that could be construed as a potential conflict of interest.

# 8 Author Contributions

# 9 Funding

This work was funded by the China Scholarship Council (201506075088 to J. M.)

# 10 Acknowledgments

The author is grateful to Prof. Kok Lay Teo at Curtin University for his support and encouragement during a 1-year visit.

## 12 Supplementary Material

Supplementary Materials and Methods

Supplementary Text

Supplementary Figures S1 to S4

Supplementary Table S1

## 13 Data Availability Statement

All data is available in the main text or in the supplementary material, and the materials required in the current study, including the schematic and PCB of the specially designed circuit for the experiment, are available from the author on reasonable request.

Supplementary Materials for

**Bridging the Gap Between Consciousness and Matter: Recurrent Out-of-Body Projection of Visual Awareness Revealed by the Law of Non-Identity**

Jinsong Meng*

*Corresponding author. E-mail: mengjinsong@uestc.edu.cn

**This PDF file includes:**

# 1 Supplementary Materials and Methods

This section includes detailed descriptions of two visual psychophysical experiments of perceiving night-shot still life under the pre-induced accommodation condition: Experiment 1, visual perception of a night-shot metal ball, and Experiment 2, visual perception of a night-shot randomly selected drawing in compliance with a single-blind procedure.

## 1.1 Participants

Eighteen volunteers participated in the experiments (14 males and four females, aged between 18 and 49 years, not colorblind), six of whom had normal vision, whereas the others had corrected-to-normal vision and had no known neurological or visual disorders. They were unaware of the specific aim of the study. The studies involving human participants were reviewed and approved by the Ethics Committee of the University of Electronic Science and Technology of China. The participants provided their written informed consent to participate in this study prior to inclusion in the experiment. The participant in Figures S3 and S4 gave written informed consent for publication of his photographs.

## 1.2 Objects to be observed

The objects to be observed included an orange metal ball and nine drawings (Figure S1). The orange ball had a diameter of 12 cm, and the drawings with black margins (L × W, 12 × 12 cm) were classified as circle, square, and triangle shapes. Each shape was available in red, green, and cyan. Each drawing was marked on the back with a unique number from 1 to 9.

## 1.3 Test equipment

The test equipment comprised a camera, an off-camera flash (Flash model DF-800II; Sidande Inc., Shenzhen, China), a pair of master-slave wireless flash triggers (Model WFC-02; Sidande Inc., Shenzhen, China) used to synchronize the actions between the camera and the off-camera flash, a specially-designed visual pre-induced accommodation circuit (VPAC) that comprised a lumen measurement circuit (LMC) and a crosshair display circuit (CDC) (Figure S2A and S2B), and a digital oscilloscope.

The LMC can measure the luminous intensity received at the center of the observed object using a light sensor (Model SFH 5711-2/3; OSRAM, Germany) with a comparable spectral sensitivity to human eyes, whereas the CDC can present a red crosshair, with a diameter of 2 cm, through eight light-emitting diodes (LEDs) after pressing the set button of the VPAC and can be automatically closed after the flash onset (Figure S2C). The digital oscilloscope, when connected with the two test points (TP) of VPAC, can sample and display the luminous intensity via the LMC. The surroundings of the crosshair of the CDC were obscured by a light barrier to prevent their light from shining on objects that were to be observed, thereby ensuring that any object that was to be observed remained unknown to the participant prior to the experiment.

Before the experiment, the experimenter linked the slave trigger to the flash, pointed the flash towards the object to be observed, and linked the two TP outputs (that is, TP1 and TP2) of VPAC to the digital oscilloscope.

**1.4 Experiment 1: Visual perception of night-shot metal ball**

First, the experimenter adhered the VPAC to the center of the ball that was fixed on the light-absorbing backdrop. The VPAC and ball were shielded by a black curtain, thereby obscuring the vision of the participant. Subsequently, one participant was seated 3 m away from the ball (Figure S3A). Thereafter, the lamp in the room was turned off and the curtain was withdrawn; the experimenter pressed the set button of the VPAC to display the crosshair and asked the participant to concentrate on the red crosshair for 5 s (Figure S3B). The experimenter then pressed the button of the master trigger and the flash shined on the ball (Figure S3C). The participant was instructed to report the shape and color of what he/she saw.

Repeated tests (n=18) showed that under the pre-induced accommodation condition, each participant correctly reported the shape and color of the metal ball after a single-pulse diffuse reflection stimulus lasting for 500 µs (Table S1).

**1.5 Experiment 2: Visual perception of night-shot randomly selected drawing**

In this experiment, each test was conducted in compliance with a single-blind procedure—a drawing was randomly selected from nine drawings, which was unknown to the participant in the dark before flash onset. For this purpose, 18 random numbers in uniform distribution (discrete) were generated in advance (none of the numbers exceeded 9 and any two adjacent numbers were different). The experiments were conducted as follows:

First, according to the first unused random number in the list, the experimenter sought out the drawing marked with the same number and fixed it on the light-absorbing backdrop. The experimenter adhered the VPAC to the center of the drawing, both of which were shielded by a curtain, thereby obscuring the vision of the participant. Subsequently, one participant was seated 3 m away from the drawing (Figure S4A). Thereafter, the lamp was turned off and the curtain was withdrawn; the experimenter pressed the set button of the VPAC and asked the participant to concentrate on the red crosshair for 5 s (Figure S4B). The experimenter then pressed the button of the master trigger and the flash shined on the drawing once (Figure S4C). The participant was instructed to report the shape and color of what he/she observed. Thus, a test task for one participant recognizing a randomly selected drawing was completed, and the random number was marked as "used."

Repeated tests (n=18) showed that under the pre-induced accommodation condition, each participant correctly reported the shape and color of a randomly selected drawing after a single-pulse diffuse reflection stimulus lasting for 500 µs (Table S1).

# 2 Supplementary Text

Although the home camera flash working at full power is safe for healthy participants, the experimenter applied a diffuser in front of the flash to soften the flash light and adjusted the output power of the flash to a lower level (here the output power was set to 1/64 of the full power). Furthermore, a dark blue filter was applied to cover the crosshair of the VPAC to attenuate the luminous intensity of the crosshair. These measures ensured that the participants were visually comfortable, and therefore, the potential negative impacts on participants in this study were even less than those of usual night photography.

# 3 Supplementary Figures

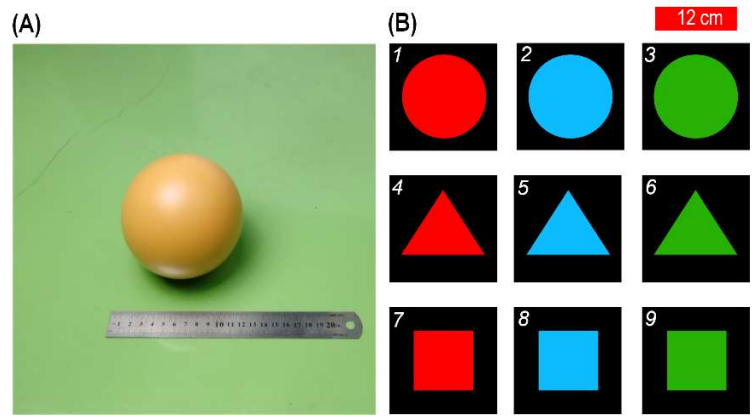

**Supplementary Figure S1.** Two types of objects to be observed (A) An orange metal ball with a diameter of 12 cm. (B) Nine drawings of different shapes in different colors with black margins, the backs of which were marked with a unique number from 1 to 9.

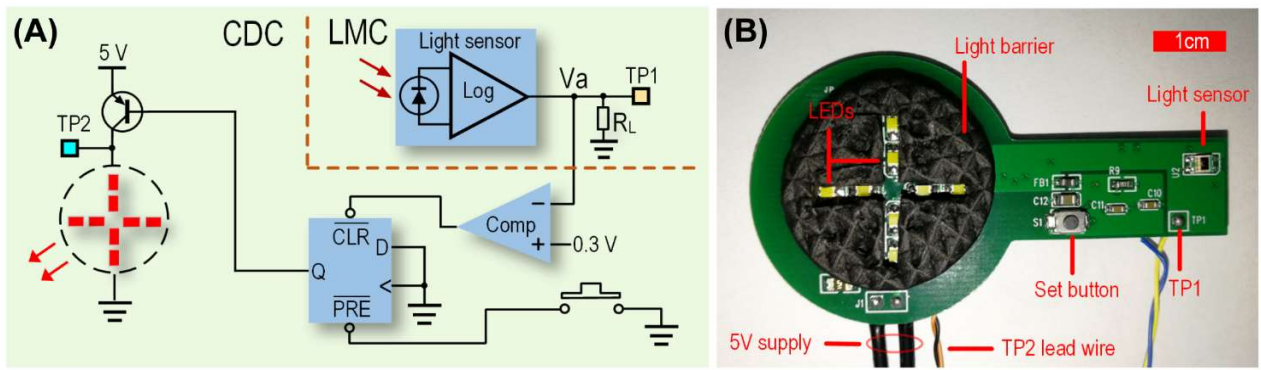

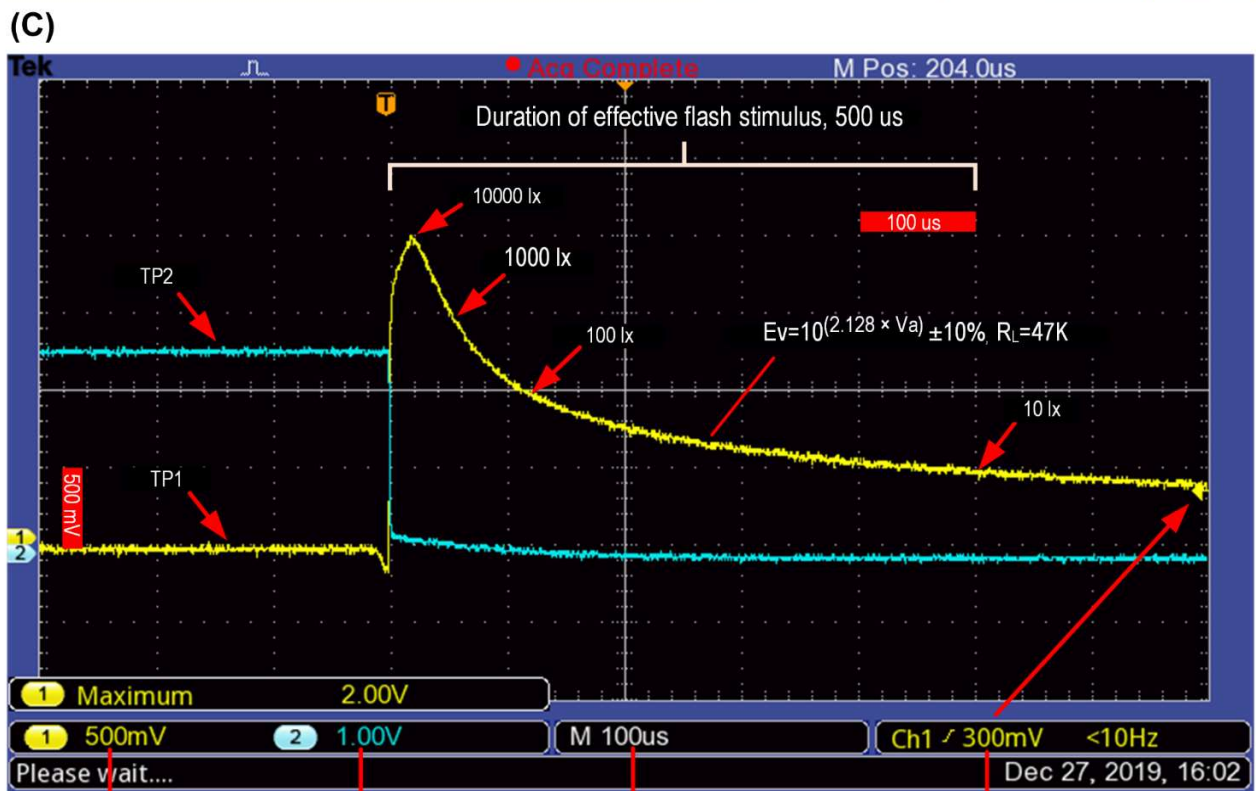

**Supplementary Figure S2.** Visual pre-induced accommodation circuit (VPAC) and its response to the flash. (A) Schematic frame of the VPAC. The module comprises two simple circuits: crosshair display circuit (CDC) and lumen measurement circuit (LMC). In the dark, on clicking the set button, CDC can present a red crosshair until the LMC receives a single-pulse flash stimulus. When Va, the output voltage of the lumen sensor, exceeds the threshold of 0.3 V, the comparator can output a low level through a flip-flop to turn off the power switch of the light-emitting diodes (LEDs). (B) Physical appearance of the VPAC. The light barrier, a trimmed, round, and self-adhesive furniture pad, with a height of 2 mm. A dark blue filter was used to cover the barrier, for attenuating the light of the LEDs (not shown). (C) Response of the VPAC to a single-pulse flash. The duration of the flash was approximately 500 μs at 1/64 full power of the flash, whereas the red crosshair went out immediately after the flash onset.

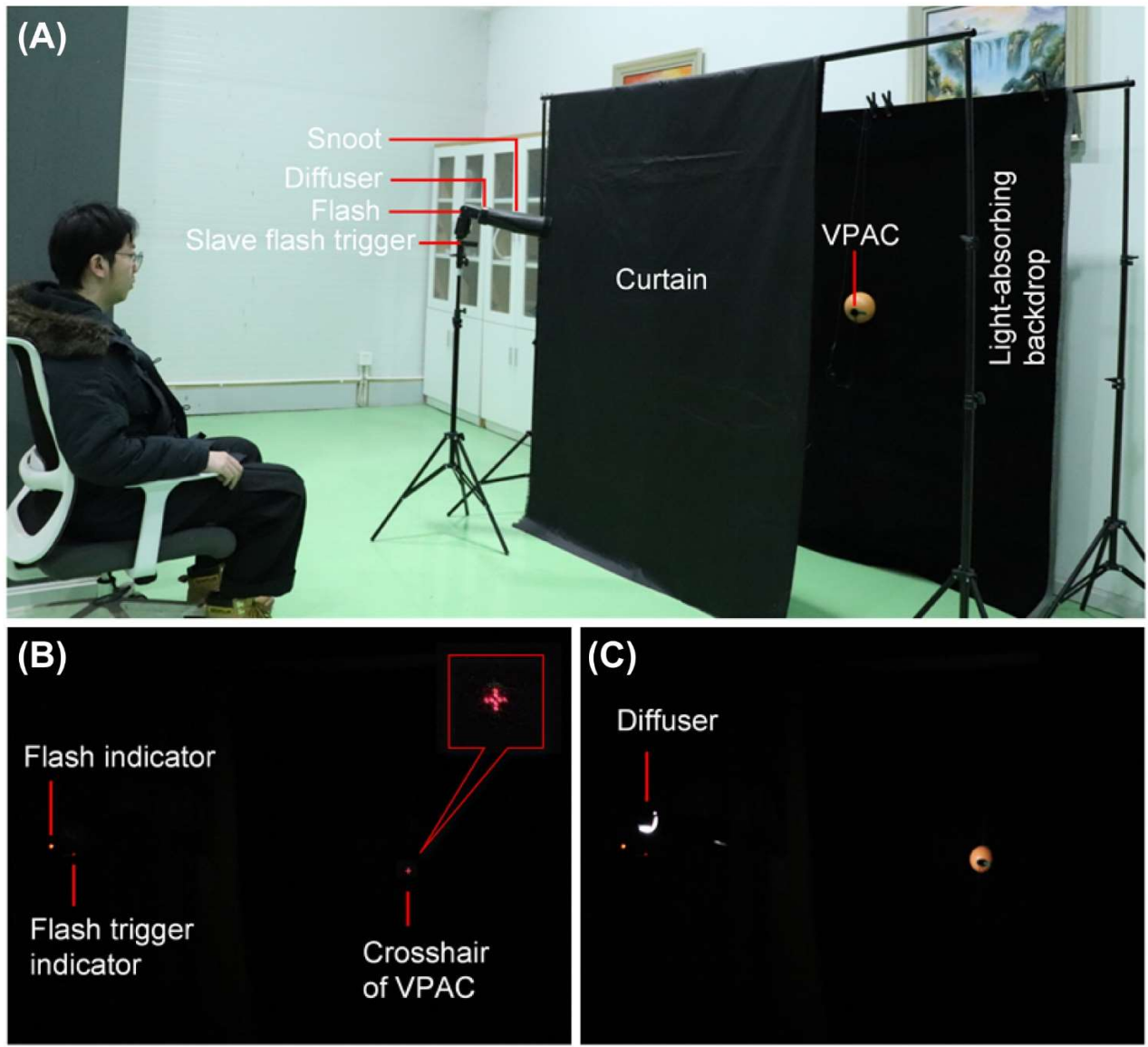

**Supplementary Figure S3.** Experimental procedure for recognizing the metal ball. (A) Layout of the participant and equipment. The participant was seated 3 m away from the ball to be observed. (B) After the curtain was withdrawn in the dark, the participant fixated on the red crosshair for 5 s. (C) After the experimenter pressed the button of the master trigger, the flash shined on the ball once

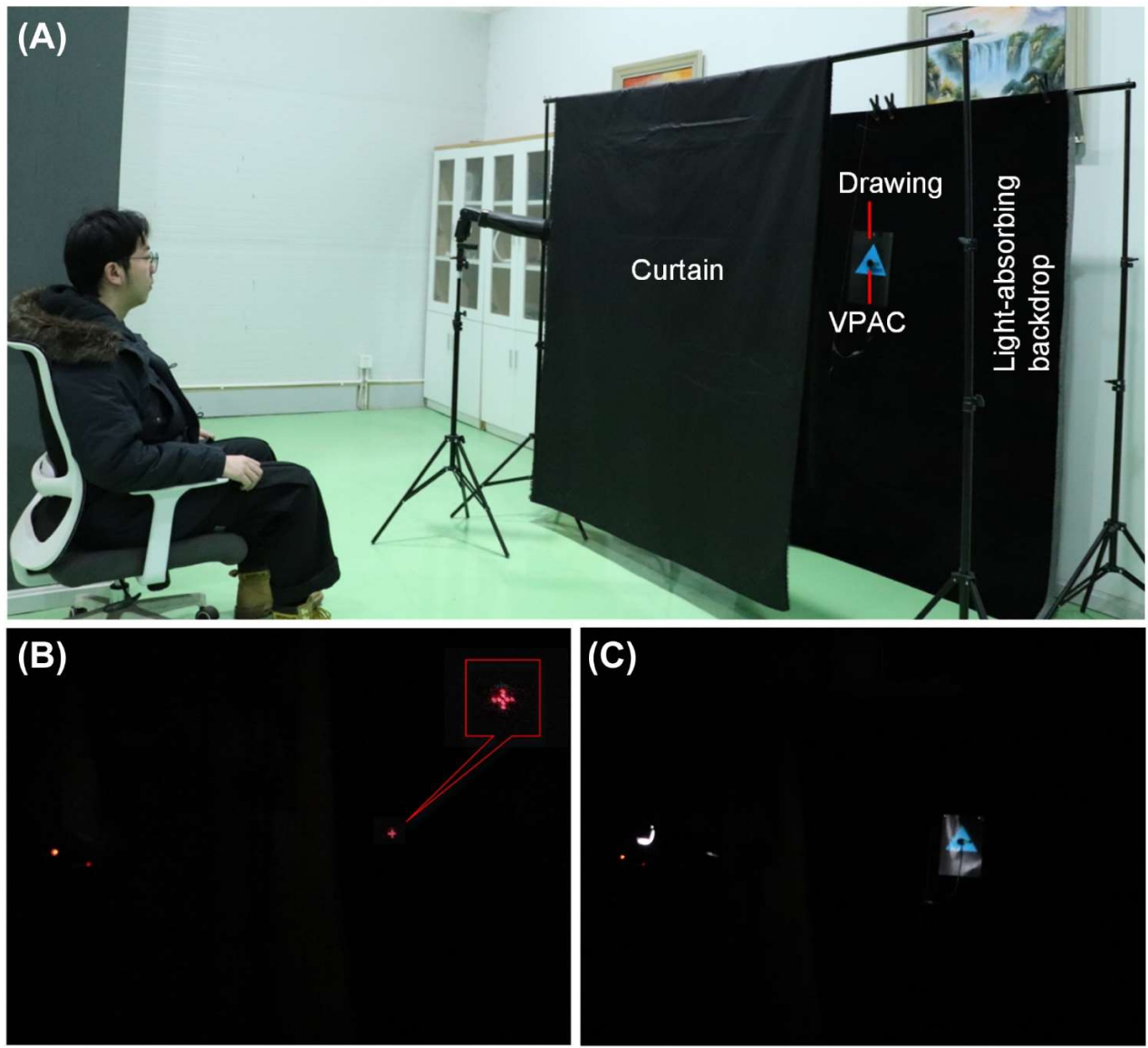

**Supplementary Figure S4**. Experimental procedure for recognizing a randomly selected drawing in compliance with a single-blind procedure. (A) Layout of the participant and equipment. The participant was seated 3 m away from the drawing to be observed. (B) After the curtain was withdrawn in the dark, the participant's gaze fixated on the red crosshair for 5 s. (C) After the experimenter pressed the button of the master trigger, the flash shined on the drawing once.

## 4 Supplementary Table

**Supplementary Table S1.** Experimental result of participants recognizing the night-shot still life

|  | Observed object | Number of participants | Stimulation intensity | Stimulation duration | Recognition rate |
|---|---|---|---|---|---|
| Experiment 1 | Metal ball | 18 | 1/64 full power | 500 µs | 100% |
| Experiment 2[a] | Drawing | 18 | 1/64 full power | 500 µs | 100% |

[a]Each test in experiment 2 was conducted in compliance with a single-blind procedure—a drawing was randomly selected from nine drawings which was unknown to the participant in the dark before flash onset.